\documentclass[12pt]{article}
\usepackage{amsmath,amsfonts,amssymb}

\begin{document}
\thispagestyle{empty}

\def\theequation{\arabic{section}.\arabic{equation}}
\def\a{\alpha}
\def\b{\beta}
\def\g{\gamma}
\def\d{\delta}
\def\dd{\rm d}
\def\e{\epsilon}
\def\ve{\varepsilon}
\def\z{\zeta}
\def\B{\mbox{\bf B}}\def\cp{\mathbb {CP}^3}

\newcommand{\h}{\hspace{0.5cm}}

\begin{titlepage}
\vspace*{1.cm}
\renewcommand{\thefootnote}{\fnsymbol{footnote}}
\begin{center}
{\Large \bf Finite-size Giant Magnons on $AdS_4 \times
CP^3_{\gamma}$ }
\end{center}
\vskip 1.2cm \centerline{\bf Changrim  Ahn and Plamen Bozhilov
\footnote{On leave from Institute for Nuclear Research and Nuclear
Energy, Bulgarian Academy of Sciences, Bulgaria.}}

\vskip 10mm

\centerline{\sl Department of Physics} \centerline{\sl Ewha Womans
University} \centerline{\sl DaeHyun 11-1, Seoul 120-750, S. Korea}
\vspace*{0.6cm} \centerline{\tt ahn@ewha.ac.kr,
bozhilov@inrne.bas.bg}

\vskip 20mm

\baselineskip 18pt

\begin{center}
{\bf Abstract}
\end{center}
\h We investigate finite-size giant magnons propagating on
$\gamma$-deformed  $AdS_4 \times CP^3_{\gamma}$ type IIA string
theory background, dual to one parameter deformation of the
$\mathcal{N}=6$ super Chern-Simoms-matter theory. Analyzing the
finite-size effect on the dispersion relation, we find that it is
modified compared to the undeformed case, acquiring $\gamma$
dependence.

\end{titlepage}
\newpage
\baselineskip 18pt

\def\nn{\nonumber}
\def\tr{{\rm tr}\,}
\def\p{\partial}
\newcommand{\non}{\nonumber}
\newcommand{\bea}{\begin{eqnarray}}
\newcommand{\eea}{\end{eqnarray}}
\newcommand{\bde}{{\bf e}}
\renewcommand{\thefootnote}{\fnsymbol{footnote}}
\newcommand{\be}{\begin{eqnarray}}
\newcommand{\ee}{\end{eqnarray}}

\vskip 0cm

\renewcommand{\thefootnote}{\arabic{footnote}}
\setcounter{footnote}{0}

\setcounter{equation}{0}
\section{Introduction}
Investigations on AdS/CFT duality \cite{AdS/CFT} for theories with
reduced or without supersymmetry are important not only
conceptually, but also for describing realistic physics. An example
of such correspondence between gauge and string theory models with
reduced supersymmetry is provided by an exactly marginal deformation
of $\mathcal{N} = 4$ super Yang-Mills theory \cite{LS95} and string
theory on a $\beta$-deformed $AdS_5\times S^5$ background suggested
in \cite{LM05}. When $\beta\equiv\gamma$ is real, the deformed
background can be obtained from $AdS_5\times S^5$ by the so-called
TsT transformation. It includes T-duality on one angle variable, a
shift of another isometry variable, then a second T-duality on the
first angle \cite{LM05,F05}.

Another interesting example is the duality between the
$\gamma$-deformed  $AdS_4 \times CP^3_{\gamma}$ type IIA string
theory and one parameter deformation of the ABJM theory \cite{ABJM},
i.e. $\mathcal{N}=6$ super Chern-Simoms-matter theory in three
dimensions. The resulting theory has $\mathcal{N}=2$ supersymmetry
and the modified superpotential is \cite{EI0808}\bea\label{dsp}
W_\gamma\propto Tr\left(e^{-i\pi\gamma/2}A_1
B_1A_2B_2-e^{i\pi\gamma/2}A_1 B_2A_2B_1\right).\eea Here the chiral
superfields $A_i$, $B_i$, $(i=1,2)$ represent the matter part of the
theory. As in the $\mathcal{N} = 4$ super Yang-Mills case, the
marginality of the deformation translates into the fact that $AdS_4$
part of the background is untouched. Taking into account that $CP^3$
has three isometric coordinates, one can consider a chain of three
TsT transformations. The result is a regular three-parameter
deformation of $AdS_4 \times CP^3$ string background, dual to a
non-supersymmetric deformation of ABJM theory, which reduces to the
supersymmetric one by putting $\gamma_1=\gamma_2=0$ and
$\gamma_3=\gamma$ \cite{EI0808}.

The dispersion relation for the giant magnon \cite{HM06} in the
$\gamma$-deformed $AdS_4 \times CP^3_{\gamma}$ background, carrying
two nonzero angular momenta, has been found in \cite{SR09}. Here we
are interested in obtaining the finite-size correction to it. To
this end, in Section 2 we introduce the $\gamma$-deformed
background, consider strings on the $R_t\times RP^3_{\gamma}$
subspace of $AdS_4 \times CP^3_{\gamma}$, and find the exact
expressions for the conserved charges and the angular differences.
In Section 3 we perform the necessary expansions, and derive the
leading corrections to the dispersion relations of giant magnons
with one and two angular momenta. In Section 4 we conclude with some
remarks.

\setcounter{equation}{0}
\section{Exact Results}
Let us first write down the deformed background. It is given by
\cite{EI0808}\footnote{There are also nontrivial dilaton and fluxes
$F_2$, $F_4$, but since the fundamental string does not interact
with them at the classical level, we do not need to know the
corresponding expressions.} \bea\nn &&
ds^2_{IIA}=R^2\left(\frac{1}{4}ds^2_{AdS_4}+ds^2_{CP^3_{\gamma}}\right),
\\ \nn && ds^2_{CP^3_{\gamma}}= d\psi^2+G\sin^2 \psi \cos^2 \psi
\left(\frac{1}{2}\cos\theta_1 d\phi_1-\frac{1}{2}\cos\theta_2
d\phi_2+d\phi_3\right)^2 \\ \nn &&+\frac{1}{4}\cos^2 \psi
\left(d\theta_1^2+G\sin^2\theta_1 d\phi_1^2\right)
+\frac{1}{4}\sin^2 \psi \left(d\theta_2^2+G\sin^2\theta_2
d\phi_2^2\right) \\ \nn && +\tilde{\gamma}G\sin^4 \psi \cos^4 \psi
\sin^2\theta_1\sin^2\theta_2 d\phi_3^2,\eea \bea \nn && B_2=-R^2
\tilde{\gamma}G\sin^2 \psi \cos^2 \psi
\\\nn &&\times\left[\frac{1}{2}\cos^2 \psi\sin^2\theta_1\cos\theta_2 d\phi_3\wedge
d\phi_1+\frac{1}{2}\sin^2 \psi\sin^2\theta_2\cos\theta_1
d\phi_3\wedge d\phi_2 \right.
\\ \nn &&
+\left.\frac{1}{4}\left(\sin^2\theta_1\sin^2\theta_2+ \cos^2
\psi\sin^2\theta_1\cos^2\theta_2 +\sin^2
\psi\sin^2\theta_2\cos^2\theta_1\right)d\phi_1\wedge d\phi_2\right]
,\eea where \bea\nn G^{-1}=1+\tilde{\gamma}^2\sin^2 \psi \cos^2 \psi
\left(\sin^2\theta_1\sin^2\theta_2+ \cos^2
\psi\sin^2\theta_1\cos^2\theta_2 +\sin^2
\psi\sin^2\theta_2\cos^2\theta_1\right).\eea The deformation
parameter $\tilde{\gamma}$ above is given by
$\tilde{\gamma}=\frac{R^2}{4}\gamma$, where $\gamma$ appears in the
dual field theory superpotential (\ref{dsp}).

\subsection{String Solutions}
In our considerations we will use conformal gauge, in which the
string Lagrangian and Virasoro constraints have the form
\bea\label{l}
&&\mathcal{L}_s=\frac{T}{2}\left(G_{00}-G_{11}+2B_{01}\right) \\
\label{00} && G_{00}+G_{11}=0,\qquad G_{01}=0.\eea Here \bea\nn
&&G_{mn} = g_{MN}\p_m X^M\p_nX^N,\h B_{mn} = b_{MN}\p_m X^M\p_nX^N,
\\ \nn &&\p_m=\p/\p\xi^m,\h m,n =
(0,1), \h(\xi^0,\xi^1)=(\tau,\sigma),\h M,N = (0,1,\ldots,9),\eea
are the fields induced on the string worldsheet.

Further on, we restrict our attention to the $R_t\times
RP^3_{\gamma}$ subspace of $AdS_4 \times CP^3_{\gamma}$, where
$\theta_1=\theta_2=\pi/2$, $\phi_3=0$, and \bea\nn &&
ds^2=R^2\left(-\frac{1}{4}dt^2+d\psi^2 + \frac{G}{4}\cos^2\psi
d\phi_1^2 +\frac{G}{4}\sin^2\psi d\phi_2^2\right),
\\ \nn && B_2 =b_{\phi_1\phi_2}d\phi_1\wedge d\phi_2
=-\frac{R^2}{4} \tilde{\gamma}G\sin^2 \psi \cos^2
\psi d\phi_1\wedge d\phi_2,
\\ \nn && G^{-1}=1+\tilde{\gamma}^2\sin^2 \psi \cos^2 \psi.\eea

To find the string solutions we are interested in, we use the ansatz
($j=1,2$) \bea\label{NRA} &&t(\tau,\sigma)=\kappa\tau,\h
\psi(\tau,\sigma)=\psi(\xi),\h
\phi_j(\tau,\sigma)=\omega_j\tau+f_j(\xi),\\ \nn
&&\xi=\alpha\sigma+\beta\tau,\h \kappa, \omega_j, \alpha,
\beta={\rm constants}.\eea Then the string Lagrangian (\ref{l}) becomes
(prime is used for $d/d\xi$) \bea\nn
&&\mathcal{L}_s=-\frac{TR^2}{2}(\alpha^2-\beta^2) \left[\psi'^2+
\frac{G}{4}\cos^2\psi\left(f'_1-\frac{\beta\omega_1}{\alpha^2-\beta^2}\right)^2
+\frac{G}{4}\sin^2\psi\left(f'_2-\frac{\beta\omega_2}{\alpha^2-\beta^2}\right)^2\right.
\\ \label{rl} &&-\left.\frac{G\alpha^2}{4(\alpha^2-\beta^2)^2}
\left(\omega_1^2\cos^2\psi+\omega_2^2\sin^2\psi\right)
+\frac{\alpha\tilde{\gamma}G}{2}\sin^2 \psi \cos^2 \psi
\frac{\omega_1 f'_2-\omega_2 f'_1}{\alpha^2-\beta^2}\right], \eea
while the constraints (\ref{00}) acquire the form \bea\nn &&
\psi'^2+
\frac{G}{4}\cos^2\psi\left(f'^2_1+\frac{2\beta\omega_1}{\alpha^2+\beta^2}f'_1+\frac{\omega_1^2}{\alpha^2+\beta^2}\right)
\\ \label{rc} &&+\frac{G}{4}\sin^2\psi\left(f'^2_2+\frac{2\beta\omega_2}{\alpha^2+\beta^2}f'_2+\frac{\omega_2^2}{\alpha^2+\beta^2}\right)
=\frac{\kappa^2/4}{\alpha^2+\beta^2},
\\ \nn && \psi'^2+
\frac{G}{4}\cos^2\psi\left(f'^2_1+\frac{\omega_1}{\beta}f'_1\right)
+\frac{G}{4}\sin^2\psi\left(f'^2_2+\frac{\omega_2}{\beta}f'_2\right)
=0 .\eea The equations of motion for $f_j(\xi)$ following from
(\ref{rl}) can be integrated once to give \bea\label{fjs}  &&
f'_1=\frac{1}{\alpha^2-\beta^2} \left[\frac{C_1}{\cos^2 \psi}
+\beta\omega_1+\tilde{\gamma}\left(\alpha\omega_2+\tilde{\gamma}C_1\right)\sin^2\psi\right],
\\ \nn && f'_2=\frac{1}{\alpha^2-\beta^2} \left[\frac{C_2}{\sin^2 \psi}
+\beta\omega_2-\tilde{\gamma}\left(\alpha\omega_1-\tilde{\gamma}C_2\right)\cos^2\psi\right]
,\eea where $C_j$ are constants. Replacing (\ref{fjs}) into
(\ref{rc}), one can rewrite the Virasoro constraints as
\bea\label{00r} &&\psi'^2=\frac{1}{4(\alpha^2-\beta^2)^2}
\Bigg[(\alpha^2+\beta^2)\kappa^2 -\frac{C_1^2}{\cos^2\psi}
-\frac{C_2^2}{\sin^2\psi}
\\ \nn &&-\left(\alpha\omega_1-\tilde{\gamma}C_2\right)^2\cos^2\psi
-\left(\alpha\omega_2+\tilde{\gamma}C_1\right)^2\sin^2\psi\Bigg],
\\ \label{01r} && \omega_1C_1+\omega_2C_2+\beta\kappa^2=0.\eea
Let us point out that (\ref{00r}) is the first integral of the
equation of motion for $\psi$. Integrating (\ref{fjs}) and
(\ref{00r}), one can find string solutions with very different
properties. All of them are related to solutions of the complex
sine-Gordon integrable model in an explicit way \cite{AB1}.
Particular examples are (dyonic) giant magnons and single-spike
strings.

\subsection{Conserved Quantities and Angular Differences}
In the case at hand, the background metric does not depend on $t$
and $\phi_j$. The corresponding conserved quantities are the string
energy $E_s$ and two angular momenta $J_j$, given by
\bea\label{gcqs} E_s=-\int d\sigma\frac{\p\mathcal{L}_s}{\p(\p_0
t)},\h J_j=\int d\sigma\frac{\p\mathcal{L}_s}{\p(\p_0\phi_j)}.\eea
On the ansatz (\ref{NRA}), $E_s$ and $J_j$ defined above take the
form \bea\label{cqs} &&E_s= \frac{TR^2}{4}\frac{\kappa}{\alpha}\int
d\xi,\\ \nn &&J_1= \frac{TR^2}{4}\frac{1}{\alpha^2-\beta^2}\int d\xi
\left[\frac{\beta}{\alpha}C_1+\left(\alpha\omega_1-\tilde{\gamma}C_2
\right)\cos^2\psi \right], \\ \nn &&J_2=
\frac{TR^2}{4}\frac{1}{\alpha^2-\beta^2}\int d\xi
\left[\frac{\beta}{\alpha}C_2+\left(\alpha\omega_2+\tilde{\gamma}C_1
\right)\sin^2\psi \right].\eea Let us remind that the relation
between the string tension $T$ and the t'Hooft coupling constant
$\lambda$ for the $\mathcal{N}=6$ super Chern-Simoms-matter theory
is given by \bea\nn TR^2=2\sqrt{2\lambda}.\eea

If we introduce the variable \bea\nn \chi=\cos^2\psi,\eea and use
(\ref{01r}), the first integral (\ref{00r}) can be rewritten as
\bea\nn \chi'^{2}
=\frac{\Omega_2^2\left(1-u^{2}\right)}{\alpha^2(1-v^2)^2}
(\chi_{p}-\chi)(\chi-\chi_{m})(\chi-\chi_{n}) ,\eea where \bea\nn
&&\chi_p+\chi_m+\chi_n=\frac{2-(1+v^2)W-u^2}{1
-u^2},\\
\label{3eqs} &&\chi_p \chi_m+\chi_p \chi_n+\chi_m
\chi_n=\frac{1-(1+v^2)W+(v W-u K)^2-K^2}{1 -u^2},\\ \nn && \chi_p
\chi_m \chi_n=- \frac{K^2}{1 -u^2},\eea and \bea\nn &&
v=-\frac{\beta}{\alpha},\h u=\frac{\Omega_1}{\Omega_2},\h
W=\left(\frac{\kappa}{\Omega_2}\right)^2,\h
K=\frac{C_1}{\alpha\Omega_2},
\\ \nn &&\Omega_1=\omega_1\left(1-\tilde{\gamma}\frac{C_2}{\alpha\omega_1}\right), \h
\Omega_2=\omega_2\left(1+\tilde{\gamma}\frac{C_1}{\alpha\omega_2}\right).\eea
We are interested in the case \bea\nn 0<\chi_{m}<\chi< \chi_{p}<1,\h
\chi_{n}<0,\eea which corresponds to the finite-size giant magnons.

In terms of the newly introduced variables, the conserved quantities
(\ref{cqs}) and the angular differences \bea\label{dad}
p_1\equiv\Delta\phi_1=\phi_1(r)-\phi_1(-r),\h
p_2\equiv\Delta\phi_2=\phi_2(r)-\phi_2(-r) ,\eea transform to
\bea\label{E} &&\mathcal{E}\equiv \frac{E_s}{TR^2}
=\frac{(1-v^2)\sqrt{W}}{\sqrt{1-u^2}}\frac{
\mathbf{K}(1-\epsilon)}{\sqrt{\chi_{p}-\chi_{n}}},
\\ \label{J1} &&\mathcal{J}_1\equiv\frac{J_1}{TR^2}=\frac{1}{\sqrt{1-u^2}}
\left[\frac{u\chi_n-v K}{\sqrt{\chi_{p}-\chi_{n}}}
\mathbf{K}(1-\epsilon) +u\sqrt{\chi_{p}-\chi_{n}}
\mathbf{E}(1-\epsilon)\right],
\\ \label{J2} &&\mathcal{J}_2\equiv\frac{J_2}{TR^2}=\frac{1}{\sqrt{1-u^2}}
\left[\frac{1-\chi_n-v\left(v W-u
K\right)}{\sqrt{\chi_{p}-\chi_{n}}}\mathbf{K}(1-\epsilon)-\sqrt{\chi_{p}-\chi_{n}}
\mathbf{E}(1-\epsilon)\right],\eea \bea
\label{p1}&&p_1=\frac{4}{\sqrt{1-u^2}}
\\ \nn &&\times\Bigg\{\frac{K}{\chi_p\sqrt{\chi_{p}-\chi_{n}}}
\Pi\left(1-\frac{\chi_{m}}{\chi_{p}}\vert 1-\epsilon\right) -
\left[uv+\tilde{\gamma}v\left(v W-u
K\right)-\tilde{\gamma}\left(1-\chi_n\right)\right]
\frac{\mathbf{K}(1-\epsilon)}{\sqrt{\chi_{p}-\chi_{n}}}
\\ \nn &&-\tilde{\gamma}
\sqrt{\chi_{p}-\chi_{n}} \mathbf{E}(1-\epsilon)\Bigg\} ,\eea \bea
\label{p2} &&p_2=\frac{4}{\sqrt{1-u^2}}
\\ \nn &&\times\Bigg\{\frac{v W-u
K}{(1-\chi_p)\sqrt{\chi_{p}-\chi_{n}}}
\Pi\left(-\frac{\chi_{p}-\chi_{m}}{1-\chi_{p}}\vert
1-\epsilon\right)-\left[v\left(1-\tilde{\gamma}K\right)+\tilde{\gamma}u\chi_{n}\right]
\frac{\mathbf{K}(1-\epsilon)}{\sqrt{\chi_{p}-\chi_{n}}}  \\
\nn &&-\tilde{\gamma}u\sqrt{\chi_{p}-\chi_{n}}
\mathbf{E}(1-\epsilon)\Bigg\},\eea where we have used the formulas
for the elliptic integrals given in Appendix A, and $\epsilon$ is
given by \bea\label{de}
\epsilon=\frac{\chi_{m}-\chi_{n}}{\chi_{p}-\chi_{n}}.\eea

From (\ref{E})-(\ref{J2}) one can see that the conserved charges are
not affected by the $\gamma$-deformation as it should be. Only the
angular differences are shifted.

Further on, we will consider the case when $\mathcal{E}$,
$\mathcal{J}_2$ and $p_1$ are large, while
$\mathcal{E}-\mathcal{J}_2$, $\mathcal{J}_1$ and $p_2$ are finite.
To this end, we will introduce appropriate expansions.

\setcounter{equation}{0}
\section{Expansions}
In order to find the leading finite-size correction to the
energy-charge relation, we have to consider the limit $\epsilon\to
0$ in (\ref{3eqs}), (\ref{E})-(\ref{J2}), and (\ref{de}). The
behavior of the complete elliptic integrals in this limit is given
in Appendix A. Taking this behavior into account, we will use the
following ansatz for the parameters $(\chi_p,\chi_m,\chi_n,v,u,W,K)$
in the solution
\bea\nn
&&\chi_p=\chi_{p0}+\left(\chi_{p1}+\chi_{p2}\log(\epsilon)\right)\epsilon,
\\ \nn &&\chi_m=\chi_{m0}+\left(\chi_{m1}+\chi_{m2}\log(\epsilon)\right)\epsilon,
\\ \nn &&\chi_n=\chi_{n0}+\left(\chi_{n1}+\chi_{n2}\log(\epsilon)\right)\epsilon,
\\
\label{Dpars} &&v=v_0+\left(v_1+v_2\log(\epsilon)\right)\epsilon, \\
\nn &&u=u_0+\left(u_1+u_2\log(\epsilon)\right)\epsilon,
\\ \nn &&W=W_0+\left(W_1+W_2\log(\epsilon)\right)\epsilon,
\\ \nn &&K=K_0+\left(K_1+K_2\log(\epsilon)\right)\epsilon .\eea
A few comments are in order. To be able to reproduce the dispersion
relation for the infinite-size giant magnons, we set \bea\label{is}
\chi_{m0}=\chi_{n0}=K_0=0,\h W_0=1.\eea 
Also to reproduce the undeformed case \cite{ABR} in the ${\tilde\gamma}\to 0$ limit,
we need to fix 
\bea\label{k2} \chi_{m2}= \chi_{n2}=W_2=K_2=0.\eea

Replacing (\ref{Dpars}) into (\ref{3eqs}) and (\ref{de}), one finds
six equations for the coefficients in the expansions of $\chi_p$,
$\chi_m$, $\chi_n$ and $W$.
They are solved by \bea\label{chi} &&\chi_{p0}=1-\frac{v_0^2}{1-u_0^2}, \\
\nn &&\chi_{p1}=
\frac{v_0}{\left(1-v_0^2\right)\left(1-u_0^2\right)(1-v_0^2-u_0^2)}
\Big\{-2v_0u_0(1-v_0^2)(1-v_0^2-u_0^2)u_1
\\ \nn &&+2\left(1-u_0^2\right)(1-v_0^2-u_0^2)
\left[K_1u_0(1+v_0^2)-(1-v_0^2)v_1\right]
\\ \nn
&&+v_0(1-v_0^2-2u_0^2)\sqrt{(1-u_0^2-v_0^2)^4-4K_1^2(1-u_0^2)^2(1-u_0^2-v_0^2)}\Big\},\eea
\bea \nn &&\chi_{p2}=
-2v_0\frac{v_2+(v_0u_2-u_0v_2)u_0}{\left(1-u_0^2\right)^2} \\
\nn &&\chi_{m1}= \frac{u_0^4
-2u_0^2(1-v_0^2)+(1-v_0^2)^2+\sqrt{(1-u_0^2-v_0^2)^4-4K_1^2(1-u_0^2)^2(1-u_0^2-v_0^2)}}
{2(1-u_0^2)(1-v_0^2-u_0^2)}, \\
\nn &&\chi_{n1}= -\frac{u_0^4
-2u_0^2(1-v_0^2)+(1-v_0^2)^2-\sqrt{(1-u_0^2-v_0^2)^4-4K_1^2(1-u_0^2)^2(1-u_0^2-v_0^2)}}
{2(1-u_0^2)(1-v_0^2-u_0^2)},
\\ \nn &&W_1=-\frac{2K_1u_0v_0(1-u_0^2)+\sqrt{(1-u_0^2-v_0^2)^4-4K_1^2(1-u_0^2)^2(1-u_0^2-v_0^2)}}
{(1-u_0^2)(1-v_0^2)}.\eea

As a next step, we impose the conditions for $\mathcal{J}_1$, $p_2$
to be independent of $\epsilon$. By expanding RHS of (\ref{J1}),
(\ref{p2}) in $\epsilon$, one gets \bea \label{j2ex}
\mathcal{J}_1=\frac{ u_0\sqrt{1-v_0^2-u_0^2}}{1-u_0^2},\eea
\bea\label{pex} p_2=2\arcsin\left(\frac{2
v_0\sqrt{1-v_0^2-u_0^2}}{1-u_0^2}\right) -4\tilde{\gamma}u_0
\frac{\sqrt{1-v_0^2-u_0^2}}{1-u_0^2},\eea along with four more
equations from the coefficients of $\epsilon$ and $\epsilon\log
\epsilon$. The equalities (\ref{j2ex}), (\ref{pex}) lead to
\bea\label{zms}
v_0=\frac{\sin\Psi}{2\sqrt{\mathcal{J}_1^2+\sin^2(\Psi/2)}},\h
u_0=\frac{\mathcal{J}_1}{\sqrt{\mathcal{J}_1^2+\sin^2(\Psi/2)}}, \h
p_2=2\left(\Psi-2\tilde{\gamma}\mathcal{J}_1\right),\eea where the
angle $\Psi$ is defined as \bea\nn \Psi=\arcsin\left(\frac{2
v_0\sqrt{1-v_0^2-u_0^2}}{1-u_0^2}\right).\eea After the replacement
of (\ref{chi}) into the remaining four equations, they can be solved
with respect to $v_1$, $v_2$, $u_1$, $u_2$, leading to the following
form of the dispersion relation in the considered approximation
\bea\label{echr1}
\mathcal{E}-\mathcal{J}_2=\frac{\sqrt{1-v_0^2-u_0^2}}{1-u_0^2}
-\frac{1}{4}\frac{\sqrt{(1-v_0^2-u_0^2)^3-4K_1^2(1-u_0^2)^2}}{1-u_0^2}
\epsilon .\eea To the leading order, the expansion for
$\mathcal{J}_2$ gives \bea\label{eps}
\epsilon=16\exp\left[-\frac{2}{1-v_0^2}\left(1-\frac{
v_0^2}{1-u_0^2}+\mathcal{J}_2\sqrt{1-v_0^2-u_0^2}\right)\right].\eea
By using (\ref{zms}) and (\ref{eps}), (\ref{echr1}) can be rewritten
as \bea\label{IEJ1} &&\mathcal{E}-\mathcal{J}_2 =
\sqrt{\mathcal{J}_1^2+\sin^2(\Psi/2)} - 4\sqrt{\frac{\sin^8(\Psi/2)}{\mathcal{J}_1^2+\sin^2(\Psi/2)}-4K_1^2}\\
\nn &&\exp\left[-\frac{2\left(\mathcal{J}_2 +
\sqrt{\mathcal{J}_1^2+\sin^2(\Psi/2)}\right)
\sqrt{\mathcal{J}_1^2+\sin^2(\Psi/2)}\sin^2(\Psi/2)}{\mathcal{J}_1^2+\sin^4(\Psi/2)}
\right].\eea

The parameter $K_1$ in (\ref{IEJ1}) can be related to the angular
difference $p_1$. To see that, let us consider the leading order in
the $\epsilon$-expansion for it: \bea\nn &&p_1=\frac{4K_1 \arctan
\sqrt{\frac{\chi_{p0}}{\chi_{m1}}-1}}{\sqrt{(1-u_0^2)\chi_{p0}\chi_{m1}(\chi_{p0}-\chi_{m1})}}
\\ \label{pt1} &&-\frac{2}{\sqrt{(1-u_0^2)\chi_{p0}}} \left[u_0v_0\log(16)
+\tilde{\gamma}\left(2\chi_{p0}-(1-v_0^2)\log(16)\right)\right]
\\ \nn &&+\frac{2}{\sqrt{(1-u_0^2)\chi_{p0}}} \left[u_0v_0
-\tilde{\gamma}(1-v_0^2)\right]\log(\epsilon).\eea So, it is natural
to introduce the angle $\Phi$ as \bea\label{ltd} \frac{\Phi}{2}=
\arctan \sqrt{\frac{\chi_{p0}}{\chi_{m1}}-1}.\eea  On the solution
for the other parameters this gives \bea\label{kn1f}
K_1=\frac{(1-v_0^2-u_0^2)^{3/2}}{2(1-u_0^2)}\sin(\Phi)
=\frac{\sin^4(\Psi/2)}{2\sqrt{\mathcal{J}_1^2+\sin^2(\Psi/2)}}\sin(\Phi)
.\eea As a result, the relation (\ref{pt1}) between the angles $p_1$
and $\Phi$ becomes \bea\label{Phi} &&\Phi=
\frac{p_1}{2}-\left(2\tilde{\gamma}-\mathcal{J}_1\frac{\sin\Psi}{\mathcal{J}_1^2+\sin^4(\Psi/2)}\right)\mathcal{J}_2+
\mathcal{J}_1\frac{\sin\Psi\sqrt{\mathcal{J}_1^2+\sin^2(\Psi/2)}}{\mathcal{J}_1^2+\sin^4(\Psi/2)}
,\eea where due to the periodicity condition we should set \bea\nn
p_1=2\pi n_1,\h n_1 \in \mathbb{Z}.\eea

Finally, in view of (\ref{kn1f}), the dispersion relation
(\ref{IEJ1}) for the dyonic giant magnons acquires the form
\bea\label{IEJ1f} &&\mathcal{E}-\mathcal{J}_2 =
\sqrt{\mathcal{J}_1^2+\sin^2(\Psi/2)} - \frac{4
\sin^4(\Psi/2)}{\sqrt{\mathcal{J}_1^2+\sin^2(\Psi/2)}}\cos\Phi\\
\nn &&\exp\left[-\frac{2\left(\mathcal{J}_2 +
\sqrt{\mathcal{J}_1^2+\sin^2(\Psi/2)}\right)
\sqrt{\mathcal{J}_1^2+\sin^2(\Psi/2)}\sin^2(\Psi/2)}{\mathcal{J}_1^2+\sin^4(\Psi/2)}
\right].\eea Based on the L\"uscher $\mu$-term formula for the
undeformed case \cite{AB5}, we propose to identify the angle
$\Psi\left(=\frac{p_2}{2}+2\tilde{\gamma}\mathcal{J}_1\right)$ with the momentum
$p$ of the magnon exitations in the dual spin chain.

Let us point out that (\ref{IEJ1f}) has the same form as the
dispersion relation for dyonic giant magnons on $R_t\times
S^3_{\gamma}$ subspace of the  $\gamma$-deformed $AdS_5\times S^5$
\cite{AB2010}\footnote{In \cite{AB2010}, we left a numerical factor
$\Lambda$ in the definition of the angle $\Phi$ undetermined.
Actually, $\Lambda= 1$. Also, the coefficient 1/2 in front of the
last term of (4.5) in \cite{AB2010} is a misprint.}. Actually, the
two energy-charge relations coincide after appropriate normalization
of the charges and after exchange of the indices 1 and 2. The only
remaining difference is in the first terms in the expressions for
the angle $\Phi$: \bea\nn &&R_t\times RP^3_{\gamma}:
\Phi=\frac{p_1}{2}+\ldots
\\ \nn &&R_t\times S^3_{\gamma}: \Phi= p_2+\ldots .\eea

All of the above results simplify a lot when one consider giant
magnons with one angular momentum, i.e. $\mathcal{J}_1=0$. In
particular, the energy-charge relation (\ref{IEJ1f}) reduces to
\bea\label{IEJ10} &&\mathcal{E}-\mathcal{J}_2 =\sin\frac{p}{2}
\left[1-4\sin^2\frac{p}{2} \cos\left(\pi
n_1-2\tilde{\gamma}\mathcal{J}_2\right)
e^{-2-2\mathcal{J}_2\csc\frac{p}{2}}\right].\eea
We want to point out that our result is different from \cite{BykFro} which has extra 
$\cos^3(p/4)$ in the denominator of the phase $\Phi$. 
\setcounter{equation}{0}
\section{Concluding Remarks}
In this letter we considered string configurations on the $R_t\times
RP^3_{\gamma}$ subspace of $AdS_4 \times CP^3_{\gamma}$. Imposing
appropriate conditions on the parameters involved, we restrict
ourselves to string solutions, which describe the finite-size giant
magnons with one and two angular momenta. Taking the limit in which
the modulus of the elliptic integrals approaches one from below, we
found the leading corrections to the dispersion relations. The
obtained results are relevant for comparison with the dual field
theory, which in the case at hand is the one parameter
$\gamma$-deformation of the $\mathcal{N}=6$ super
Chern-Simoms-matter theory in three space-time dimensions.

It would be interesting to understand how to reproduce the
dispersion relation (\ref{IEJ1f}) by using L\"uscher��s approach
\cite{L86}. The dispersion relation has a specific
$\tilde{\gamma}$-dependence for finite $\mathcal{J}_2$, and it is
not quite clear how such a dependence follows from the $S$-matrix
approach. To this end, we need a generalization of the L\"uscher's
formulas for the case of nontrivial {\it twisted} boundary
conditions.

\section*{Acknowledgements}
This work was supported in part by WCU Grant No. R32-2008-000-101300
(C. A.) and DO 02-257 (P. B.).

\def\theequation{A.\arabic{equation}}
\setcounter{equation}{0}
\begin{appendix}

\section{Elliptic Integrals and $\epsilon$-Expansions}
The elliptic integrals appearing in the main text are given by
\bea\nn &&\int_{\chi_{m}}^{\chi_{p}}
\frac{d\chi}{\sqrt{(\chi_{p}-\chi)(\chi-\chi_{m})(\chi-\chi_{n})}}
=\frac{2}{\sqrt{\chi_{p}-\chi_{n}}}\mathbf{K}(1-\epsilon),\\ \nn
&&\int_{\chi_{m}}^{\chi_{p}} \frac{\chi
d\chi}{\sqrt{(\chi_{p}-\chi)(\chi-\chi_{m})(\chi-\chi_{n})}}
\\ \nn &&
=\frac{2\chi_{n}}{\sqrt{\chi_{p}-\chi_{n}}}\mathbf{K}(1-\epsilon)+2\sqrt{\chi_{p}-\chi_{n}}
\mathbf{E}(1-\epsilon),
\\ \nn &&\int_{\chi_{m}}^{\chi_{p}}
\frac{d\chi}{\chi\sqrt{(\chi_{p}-\chi)(\chi-\chi_{m})(\chi-\chi_{n})}}
=\frac{2}{\chi_{p}\sqrt{\chi_{p}-\chi_{n}}}
\Pi\left(1-\frac{\chi_{m}}{\chi_{p}}\vert 1-\epsilon\right),
\\ \nn&&\int_{\chi_{m}}^{\chi_{p}}
\frac{d\chi}{\left(1-\chi\right)\sqrt{(\chi_{p}-\chi)(\chi-\chi_{m})(\chi-\chi_{n})}}
\\ \nn &&=\frac{2}{\left(1-\chi_{p}\right)\sqrt{\chi_{p}-\chi_{n}}}
\Pi\left(-\frac{\chi_{p}-\chi_{m}}{1-\chi_{p}}\vert
1-\epsilon\right),\eea where \bea\nn
\epsilon=\frac{\chi_{m}-\chi_{n}}{\chi_{p}-\chi_{n}}.\eea

We use the following expansions for the complete elliptic integrals
\cite{w} \bea\nn &&\mathbf{K}(1-\epsilon)=
-\frac{1}{2}\log\left(\frac{\epsilon}{16}\right)-\frac{1}{4}\left(1+\frac{1}{2}\log\left(\frac{\epsilon}{16}\right)\right)
\epsilon +\ldots,
\\ \nn
&&\mathbf{E}(1-\epsilon)=
1-\frac{1}{4}\left(1+\log\left(\frac{\epsilon}{16}\right)\right)\epsilon
+\ldots,
\\ \nn &&\Pi(-n|1-\epsilon)=\frac{2\sqrt{n}\arctan(\sqrt{n})-\log\left(\frac{\epsilon}{16}\right)}{2(1+n)}
\\ \nn
&&-\frac{2-4\sqrt{n}\arctan(\sqrt{n})+(1-n)\log\left(\frac{\epsilon}{16}\right)}{8(1+n)^2}\epsilon
+\ldots, \h n>0.\eea

We use also the equality \cite{PBM-III} \bea\nn
\Pi(\nu|m)=\frac{q_{1}}{q}\Pi(\nu_1|m)-\frac{m}{q\sqrt{-\nu\nu_1}}\mathbf{K}(m),\eea
where \bea\nn &&q=\sqrt{(1-\nu)\left(1-\frac{m}{\nu}\right)},\h
q_1=\sqrt{(1-\nu_1)\left(1-\frac{m}{\nu_1}\right)},\\ \nn
&&\nu=\frac{\nu_1-m}{\nu_1-1},\h \nu_1<0,\h m<\nu<1 .\eea

\end{appendix}

\end{document}